# AstroAtom: How to Make an Atomic Blog in Your Own Kitchen
**Summary of the Workshop:** *Uncertainties in Atomic Data and How They Propagate in Chemical Abundances*

Valentina Luridiana[1,2], Jorge García-Rojas[1,2], Kanti Aggarwal[3], Manuel Bautista[4], Maria Bergemann[5], Franck Delahaye[6], Giulio del Zanna[7], Gary Ferland[8], Karin Lind[5], Arturo Manchado[1,2], Claudio Mendoza[9], Adal Mesa Delgado[1,2], Manuel Núñez Díaz[1,2], Richard A. Shaw[10], and Roger Wesson[11]

[1]*Instituto de Astrofísica de Canarias, Vía Láctea s/n, La Laguna, E-38205, Spain*

[2]*Departamento de Astrofísica, Universidad de La Laguna, La Laguna, E-38205, Spain*

[3]*Astrophysics Research Centre, School of Mathematics and Physics, Queen's University Belfast, Belfast BT7 1NN, Northern Ireland, UK*

[4]*Department of Physics, Western Michigan University, Kalamazoo, MI 49008, USA*

[5]*Max-Planck Institute for Astrophysics, Karl-Schwarzschild Str. 1, 85741, Garching, Germany*

[6]*Observatoire de Paris, 5, Place Jules Janssen, 92190 Meudon, France*

[7]*Dept. of Applied Mathematics and Theoretical Physics, CMS, University of Cambridge*

[8]*University of Kentucky, Lexington, KY, USA*

[9]*Centro de Física, Instituto Venezolano de Investigaciones Científicas, 20632 Caracas, Venezuela*

[10]*NOAO, Tucson, AZ 85719, USA*

[11]*Department of Physics and Astronomy, University College London, Gower Street, London WC1E 6BT, UK*

**Abstract.**
This workshop brought together scientists (including atomic physicists, theoretical astrophysicists and astronomers) concerned with the completeness and accuracy of atomic data for astrophysical applications. The topics covered in the workshop included the evaluation of uncertainties in atomic data, the propagation of such uncertainties in chemical abundances, and the feedback between observations and calculations. On a different level, we also discussed communication issues such as how to ensure that atomic data are correctly understood and used, and which forum is the best one for a fluid interaction between all communities involved in the production and use of atomic data. This paper reports on the discussions held during the workshop and introduces

> *AstroAtom*, a blog created as a platform for timely and open discussions on the needs and concerns over atomic data, and their effects on astronomical research.

## 1.  Introduction

Atomic data, such as energy levels, radiative transition probabilities and collisional excitation rates, are a necessary ingredient in the modelling of astrophysical plasmas. When choosing which data to use for a particular calculation, astrophysicists need to assess which particular dataset is the most reliable and decide how to fill in gaps in selected datasets. Additionally, they often wish to assess the uncertainties in the chosen data in order to know how they will affect the accuracy of their final results. These are formidable tasks for those lacking a deep knowledge of atomic physics, and yet results may depend strongly on such choices. The workshop *Uncertainties in Atomic Data and How They Propagate in Chemical Abundances*, held in Tenerife (Spain) during 25—27 October 2010, was devoted to explore these issues and their solutions. In this paper we report on the discussions held and the main conclusions that were drawn during the workshop. Since many of these issues can be mitigated, if not solved, by a more fluid communication between experts in the various fields involved, the workshop wrapped up with a commitment to put in place a mechanism to favour such communication.

## 2.  Do Uncertainties Matter?

The answer is certainly yes! Many important astrophysical questions, including the synthesis of primordial He, stellar evolution models, stellar atmosphere models, yields for supernovae of various types, etc., depend upon accurate atomic data. More directly, the interpretation of comparisons between theory and observations depends critically on thorough assessments of the uncertainties in chemical abundances that are derived from atomic data. Accurately quantifying the effect of uncertainties is often difficult because of the complex ways in which different types of atomic data affect the calculations. For example, in computing elemental abundances from emission lines in a spectrum one has, on the one hand, the effects of uncertainties from collisional excitation rates and transitions probabilities on the derived ionic abundances, and, on the other, the effects of photoionization and recombination rates on the ionization correction factors.

### 2.1.  How Large Are Typical Uncertainties?

The point of computing accurate atomic data is to enable reliable determinations to be made of chemical abundances in a variety of physical contexts. In order to achieve the necessary level of accuracy to address important scientific questions a proper accounting of the contributions from all sources of uncertainty is essential. A recent study of chemical abundances derived for a common sample of ionized nebulae by multiple authors showed surprisingly large discrepancies: from ~40% to as much as a factor of a few. The analysis showed that a number of factors were responsible, including problems with observing technique, imprecise calibration, differing analysis methods, wavelength coverage of the critical transitions and spectral resolution. Uncertainties in the atomic data undoubtedly also contributed, though they are often difficult to quantify. With care, observational uncertainties can be kept below 10%.

For atomic data there is no such thing as a "typical" magnitude of uncertainty because uncertainties are governed by many factors. Generally speaking, uncertainties in energy levels are smaller than uncertainties in transition probabilities, which are in turn smaller than those in collisional rates (see also Bautista 2011 for more details). The uncertainties in collisional data for weak transitions are larger than those in data for strong transitions. Near energy thresholds, collision strengths are often dominated by resonances, and the accuracy of a calculation depends on the exact position (apart from width and magnitude) of these resonances, which is difficult to pin down theoretically. For the same reason, the accuracy cannot be the same in all temperature ranges, as not all the energy ranges are equally affected by resonances. The same effect is observed in dielectronic recombination. All in all, for a typical ion and typical conditions of interest in nebular astrophysics, the uncertainty in transitional probabilities for strong dipole allowed transitions are $\sim 10\%$, in transition probabilities for forbidden transitions $\sim 30\%$, and in (effective) collision strengths $\sim 30-50\%$ (but bear in mind that these are just rough guidelines and should not be intepreted as representative of any specific ion). In addition to the data themselves, users should ideally be provided with a quantitative estimate of the accuracy associated with them (in way similar to the NIST compilation). Alas, since there are no standardized ways to quantify uncertainties on atomic calculations, one will not often find them quoted in a paper, and users have to make their own assessment of the accuracy of the data they use. How? The simplest solution is always to ask either the author or other specialists who might be able to give, if not a formal error estimate, at least an informed opinion on the data. Other users may also be helpful by providing information on which data they use, why, and whether they get consistent results with them.

An indication of the accuracy of a given dataset can be obtained by comparing it to other calculations, to laboratory measurements or to observational data. In the first case, the dataset is compared to previously published data: consistency among them suggests that the results have converged, while lack of consistency can be interpreted as an indication of residual uncertainty, depending on how optimistic one is in his assessment of new data (which are generally, but not always, more reliable than the old ones; see, e.g., the discussion in Aggarwal 2011). Unfortunately, consistency may also mean that both calculations suffer from a common, unknown bias. The most useful papers discuss the differences between old and new data determinations, but regrettably this is not always the case.

Few laboratory data are available to compare against calculations due to the technical difficulty and the high cost of producing them. Energy levels are almost the only kind of data for which obtaining laboratory data is feasible. Measurements of collisional cross sections at a few energies may also be helpful for validating theory, but they may be intrinsically uncertain and suffer from the obvious limitation of not covering an entire energy range. It would certainly be desirable if more support were given for laboratory measurements to improve on this aspect since, for many quantities, laboratory measurements are still insufficient to provide enough estimates on the uncertainties.

Finally, theoretical data can be benchmarked against observations to provide estimates of uncertainties; real-life examples are the comparison between the observed and predicted [O III] 4959/5007 ratio, the [S II] temperature, or the intensities of the [Ne V] infrared lines in nebulae (see e.g. the talks by R. A. Shaw and C. Mendoza). For the XUV, comparisons between observed and predicted line intensities such as those made in the benchmark work of Del Zanna (2011) are also useful. During the work-shop, we identified the need for high-quality spectral observations, from the X-ray to the infrared, to properly benchmark atomic data. Of course, observed line intensities are subject to their own observational errors, and in many cases the comparison also depends on modelling assumptions.

## 2.2. How Can Uncertainties Be Reduced?

Reducing uncertainties is a path that involves computing better and more extensive atomic calculations and benchmarking the results of these against experimental determinations and astronomical observations. This is too often an arduous process that requires effective interaction among all participants. One of the critical factors in improving the uncertainties is the adoption by the astronomical community of the "best" atomic data, however defined. In practice there are only a modest number of tools that astronomers use for abundance calculations, and the curators for some of them (Cloudy, Chianti, IRAF/nebular) have participated in this conference. Keeping these tools up to date is one way to promote the use of the best available data. Another useful approach is for the authors of these tools to compare their calculations on reference datasets on a regular basis. It is also important for the curators of these tools both to make it easy for users to see what atomic data are being used and to cite the primary sources of the data in their papers. The process also needs technical and human resources, which are often limited by lack of proper funding and job opportunities for scientists in the field, to the point where maintaining expertise through future generations seems to be in jeopardy.

## 3. How Can Communication be Improved?

Fluent interaction between astrophysicists and atomic physicists has immediate benefits to all parties. It helps astrophysicists in deciding among different sources of atomic data and their uncertainties. It benefits producers of atomic data in helping them identify the most pressing user needs as well as opportunities for benchmarking the atomic data. An example of this is the following shopping list of urgently-needed atomic data, which was assembled during the workshop:

- Photoionization cross sections of neutral and singly ionized iron and trans-iron species

- Revision of the Einstein A-values for the allowed transitions of S II and S III

- Dielectronic recombination of S II and S III

- High-n collision strengths for Li and Na

- Revision of the Ne V infrared diagnostics in nebulae

- Atomic data for low-ionization (singly, doubly and triply) *s*-process elements: Se, Br, Kr, Rb, Xe, Ba and Pb

We believe that this kind of communication between data users and producers is key to making the most of the existing data. In the workshop, we have considered several alternative formats for such a forum and finally decided to open a blog devoted to the discussion of atomic data and their astrophysical applications at the web address

http://astroatom.wordpress.com/. We hope that our blog will soon become both a communication channel and a repository of handy information on atomic data. If you are interested in participating, you are welcome to register and send your contribution.

**Acknowledgments.** We thank the Instituto de Astrofísica de Canarias for hosting and sponsoring this meeting. The meeting has been partly funded by the Spanish Ministry of Science and Innovation through grant AYA2010-11205-E.